\documentclass[aps,draft,epsf,twocolumn,superscriptaddress,eqsecnum]{revtex4}
\input epsf
\newcommand{\be}{\begin{equation}}
\newcommand{\ee}{\end{equation}}
\newcommand{\bea}{\begin{eqnarray}}
\newcommand{\eea}{\end{eqnarray}}

\begin{document}

\title{Theory of the Quantum Hall Smectic Phase, I: Low Energy Properties of
the Quantum Hall Smectic Fixed Point}
\author{Daniel G.\ Barci}
\affiliation{Department of Physics, University of Illinois at
Urbana-Champaign, 1110
W.\ Green St.\ , Urbana, IL  61801-3080, USA}
\affiliation{Departamento de F\'\i sica Te\'orica,
Universidade do Estado do Rio de Janeiro,Rua S\~ao Francisco Xavier 524, 20550-
013, 
Rio de Janeiro, RJ, Brazil}
\altaffiliation[Permanent Address: ]{Departamento de F\'\i sica Te\'orica,
Universidade do Estado do Rio de Janeiro, Rua S\~ao Francisco Xavier 524, 20550-
013, 
Rio de Janeiro, RJ, Brazil}
\author{Eduardo Fradkin}
\affiliation{Department of Physics, University of Illinois at
Urbana-Champaign, 1110
W.\ Green St.\ , Urbana, IL  61801-3080, USA}
\author{ Steven A.\ Kivelson}
\affiliation{Department of Physics,
U.\ C.\ L.\ A.\, Los Angeles, CA 90095}
\author{Vadim Oganesyan}
\affiliation{Department of Physics,
U.\ C.\ L.\ A.\, Los Angeles, CA 90095}
\date{\today}



\begin{abstract}
We develop an effective low energy theory of the Quantum Hall (QH) Smectic or stripe phase of a
two-dimensional electron gas in a large magnetic field in terms of its
Goldstone modes and of the charge fluctuations on each stripe.  
This liquid crystal phase corresponds to a fixed point which is
explicitly demonstrated to be  stable against 
quantum fluctuations at long wavelengths. 
This fixed point theory also allows an unambiguous
reconstruction of the electron operator. We find that 
quantum fluctuations are so severe that the electron
Green function decays faster than any power-law, although slower than
exponentially, and that consequently there is a deep pseudo-gap in the 
quasiparticle spectrum. We discuss, but do not
resolve the stability of the quantum Hall smectic to
crystallization.  Finally, the role of Coulomb interactions
and the low temperature thermodynamics of the QH smectic
state are analyzed.
\end{abstract}
\maketitle






Recent experiments on extremely high mobility 
two-dimensional electron gases (2DEG) in large magnetic 
fields by 
M.\ Lilly and coworkers\cite{Lilly} and by R.\ Du and coworkers\cite{du}, have 
revealed an
unusually large and strongly temperature dependent anisotropy in the transport 
properties. These, and subsequent experiments\cite{subsequent}, have made 
it clear
that this anisotropy is an intrinsic property of a new, anisotropic metallic
phase of the 2DEG. 
The anisotropic metal, rather than the fractional quantum Hall effect, 
apparently
dominates the physics of all partially filled Landau levels (LL) with LL index 
$N \geq 2$.

Motivated by these experiments and theoretical work on Hartree-Fock states with 
stripe
order\cite{platzman,fogler,chalker,tudor}, and exploiting an
analogy with the stripe related phases of other strongly correlated electron 
systems\cite{nature},
two of us  proposed\cite{FK} that the ground states of Quantum Hall 
Systems with 
partially filled Landau levels with $N \ge 2$ are predominantly
electronic liquid crystalline. These phases, with broken rotation and sometimes translation symmetry,
are intermediate 
between the isotropic quantum fluid and the quasi-classical electronic crystal.
In particular, we argued for the existence of quantum  Hall smectic, nematic, 
and hexatic 
phases.

The simplest liquid crystalline phase to visualize is the smectic or stripe 
ordered phase,
which breaks translation symmetry in one direction, but is none-the-less a fluid 
in the sense that
there is no gap to current carrying states and there is a non-integer number of 
electrons per 
magnetic unit cell.  
As mentioned above,  early Hartree-Fock calculations suggested that the ground 
state of
the 2DEG in large magnetic fields is a unidirectional charge density wave 
(CDW)\cite{platzman}.
These results are manifestly incorrect in the lowest 
Landau level, where the fractional quantum Hall effect occurs,
and the ground state is, to good approximation, the Laughlin 
state\cite{laughlin}.
But, for a high enough
Landau level index $N$,
a stripe phase can be shown to be  a 
reasonable ground state of the 2DEG for filling factors
close to half-filling of the partially filled Landau 
level\cite{fogler,chalker,tudor}.  Here, the stripes are the
result of a
competition between  {\em effective attractive short-range forces} and the 
familiar (long range) 
Coulomb interactions. 
We will see below that these mean field pictures of the stripe state are a 
good starting 
point for a description of 
a quantum Hall smectic.
However, at Hartree-Fock level, the smectic phase can easily be 
seen\cite{FK,fertig,MF} to
be unstable to crystallization, that is to the occurrence of  a charge-density 
wave along
the stripe direction which breaks translational symmetry and produces an 
insulating state
which can be thought of as an anisotropic Wigner crystal.   

In the present paper, we will 
develop a theory of the unpinned 
quantum Hall smectic phase, and investigate its low
energy properties. Based on our earlier work\cite{FK,FKMN}, it is our
belief that it is the quantum Hall nematic state, not the smectic, that
is the most experimentally important phase. However, these are new phases
of matter, and their precise characterization  is 
important 
in its own right, and for possible relevance to future experiments. Moreover,
a mean-field theory of the quantum Hall nematic, 
to serve as a starting point for a similar analysis, has not yet
been developed although important progress in this direction has been 
made\cite{nematic}.
Serious progress has also been made towards understanding the
nematic Fermi fluid in zero external magnetic field\cite{vadim}

The central result of this paper is an effective low energy
theory for the QH smectic which can be expressed in terms of two canonically
conjugate sets of degrees of freedom: the
displacement fields of the stripes, {\it i.\ e.\/} the Goldstone mode of the
spontaneously broken translational symmetry, and the chiral edge modes of each
stripe. In a separate publication\cite{paper2} two of us we will give a microscopic
derivation of this effective low energy theory. 

The quantum smectic is also interesting from a broader conceptual 
point of view, in that the Goldstone modes are so soft that coupling 
to them destroys all coherent single-electron motion, even at 
temperature $T=0$.  Indeed, we find that the electron Green function,
$G(x,y,t)$,  is highly anisotropic.  It falls off more slowly than an
exponential, but more rapidly than any power law as a function of time,
$t$, or distance along a stripe,
$x$; it falls at least exponentially as a function of displacement
perpendicular to the stripes, $y$.  For instance, for zero spatial separation,
$G(0,0,t)\sim \exp[ - A\log^2(t/t_0) ]$, which implies a strong pseudo-gap
in the local density of states.  Despite this, for a system with short range
interactions, the specific heat at low $T$,
due to the soft Goldstone modes, is $C_V\sim T\log(T_0/T)$ - there are even more
low energy modes than in a Fermi liquid! For the case of Coulomb
interactions we find instead a $T$ linear law.

\begin{figure}
\begin{center}
\leavevmode
\noindent
\epsfxsize=7 cm
\epsfysize=7 cm
\epsfbox{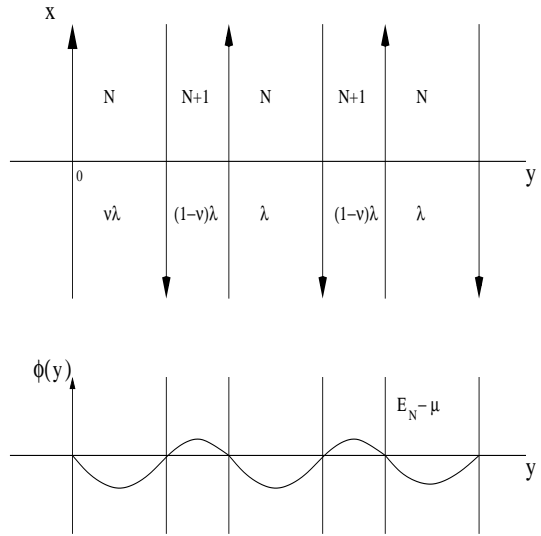}
\end{center}
\caption
{Schematic representation of the stripe solution. Top panel: the chiral edge states of each
stripe; here $N$ and $N+1$ label the number of filled Landau levels in each region,
$\lambda$ is the period of the stripe (wavelength), and $\nu$ is the effective filling factor of
the partially filled Landau level. Bottom panel: effective potential $\phi(y)$ for
the stripe solution; $y$ is the coordinate perpendicular to the stripe, $\mu$ is the chemical
potential and $E_N$ is the energy of Landau level $N$.}
\label{figstripe}
\end{figure}

A smectic state (quantum or classical) is a state with spontaneously broken 
continuous symmetries\cite{deGennes,lubensky}.  In the zeroth order 
description, given by the Hartree-Fock theory, 
the sample is spontaneously divided into strips with filling 
factors that alternate
between two successive integers, $N$ and $N+1$, as shown in Fig.\
\ref{figstripe}. 
At this level, the high density regions are separated
from the low density regions by (straight) edge states with alternating 
chirality. As the filling factor
of the partially filled level changes the width and period of these 
strips changes,
which implies that this state is compressible. As a consequence, the Hall
conductivity varies linearly as the filling factor varies across the partially
filled Landau level. 

It is tempting to think of the smectic as an array of chiral, one 
dimensional wires (internal edge states), interacting by some (possibly complicated) 
effective interaction induced by the fact that the edges are 
self-consistently generated, and hence fluctuating.  This approach 
was advocated in some of our recent papers\cite{FK,EFKL}. It was also advocated  
in an insightful paper by MacDonald and Fisher\cite{MF} who proposed an 
effective theory for a system of
coupled chiral Luttinger liquids compatible with rotational invariance, 
as required for a true smectic
state. In fact, the theory of MacDonald and Fisher is essentially 
equivalent to the theory for the
quantum hall smectic that we introduce and discuss in this paper. 

However, we find that the picture of
the quantum Hall smectic as an array of coupled Luttinger liquids in 
fact can 
obscure the underlying physics of this state.
This picture suggests that some sort of quasi-one-dimensional low-energy 
theory can explain the
physics of this state and in particular that it can be used to analyze its stability. 
Thus, on the basis of this analysis, it was argued in ref.\ \cite{FK} and \cite{MF}
that even beyond Hartree-Fock theory the smectic phase would always unstable 
to crystallization.  
While this approach is reasonable in the case in which the internal edges are 
pinned by an external, 
periodic potential, we will see, below, that this stability analysis based 
on a $1+1$-dimensional
scaling may not be safe in the absence of explicit symmetry breaking. 
The physical reason is  that the Goldstone modes of the smectic are so soft that 
they dominate the low energy physics, and the resulting fixed point 
is a strongly anisotropic $2+1$-dimensional system. 

In fact, 
H.\ A.\ Fertig and coworkers\cite{fertig,fertig-cote,fertig-stability}, have 
carried out extensive numerical time-dependent 
Hartree-Fock calculations, and showed that their numerical results can be described 
qualitatively by a quantum smectic with essentially the same properties that we find here. 
In particular, they find that the unpinned quantum Hall smectic state 
can be stable. In addition, C{\^o}t{\'e} and Fertig have analyzed their time-dependent 
Hartree-Fock results in terms of an effective elastic theory similar to ours.
Recently, Fogler and Vinokur studied the classical
hydrodynamics at finite temperature of the Hall smectic\cite{fogler-vinokur}. 

These
conflicting results, and our earlier arguments\cite{nature}  of a first order
transition for the smectic-crystal transition in the absence of pinning and of a second
order transition\cite{EFKL} in the presence of pinning, show that 
the question of the stability of the quantum Hall smectic to crystallization is still an
open problem. We will not resolve the issue  of the stability of te QH smectic against crystallization in this paper, as it turns out
to be very subtle, but we will discuss the present status of
this problem. 

While this paper was being completed we became aware of the recent work by A.\
Lopatnikova and coworkers\cite{bert} who have derived an
effective low energy theory for the quantum Hall smectic which is essentially 
equivalent to the one we present in this paper.
Although, for the most part, the
results of the present paper, as well as those of our microscopic
theory which will be discussed elsewhere\cite{paper2}, 
are in  agreement with the results of ref.\
\cite{bert}, there are
differences in the analysis of the
consequences of the effective low energy theory. 

This paper is organized as follows: In Section \ref{sec:phenomenology},
we present a simple, phenomenological derivation of the quantum Hall 
smectic fixed point Hamiltonian.  In Section
\ref{sec:CF} we use this effective field theory to compute the correlation
functions of the quantum Hall smectic, including the electron propagator. 
Since the fixed-point Hamiltonian is formally scale invariant, one
generally expects these correlation functions to be power-laws, with
exponents determined  by the scaling dimension of the fields;  this
expectation is met, in large measure, but there are various logarithms
that appear in certain limits which violate scaling.  This is the origin
of the subtleties in the stability analysis.
  In Section
\ref{sec:coulomb} we repeat the previous calculations in the presence of
unscreened Coulomb interactions. In Section
\ref{sec:thermo} we discuss the low temperature thermodynamic properties
of the QH smectic and in particular we compute the specific heat. In
section
\ref{sec:stability} we discuss the issue of the stability of the quantum
Hall smectic towards crystallization.  Finally, in Section
\ref{conclusions} we  present our conclusions. 

\section{Effective Field Theory}
\label{sec:phenomenology}

Consider a stripe crystal in a large magnetic field.  In the classical ground-
state, there is a charge density wave with fixed
wave-length running along each stripe and the stripes
(labeled by an index $j$) are spaced by distance $\lambda$ and are straight and 
parallel; without loss of generality they can be taken to run in the $\hat x$ direction.  
Smooth deformations of this state can be described in terms of the local 
displacement of the charge-density wave (CDW) along the stripe direction,
$\phi_j(x)$, and the transverse displacement of the stripe from its classic 
ground state position,
$u_j(x)$.  In terms of these variables, the action describing the dynamics of 
this system is
\bea
S=&& \lambda \sum_j\int dt dx\left\{ {\cal L}_{sm}+{\cal L}_{lock} + {\cal
L}_{irr}\right\} 
\\
{\cal L}_{sm}= &&  \frac {eB} {\lambda}  u_j \partial_t \phi_j - \frac 
{\kappa_{\parallel}}
2  (\partial_x\phi_j)^2  
\nonumber \\
& & \qquad \mbox{}
-\frac {\kappa_{\perp}} 2 \left ( \frac {u_j-u_{j+1}} \lambda\right )^2 -\frac Q 
2
(\partial_x^2 u_j)^2
 \\
{\cal L}_{lock} = && V\cos[\alpha(\phi_j-\phi_{j+1})+\beta \partial_x u] \\
{\cal L}_{irr}= && \frac 1 2 [M_1(\dot u_j)^2 + M_2(\dot \phi_j)^2] + 
\gamma (\partial_x\phi_j)(\partial_xu_j)^2 + \ldots .\nonumber
\eea
where we have used that the current on the $j$-th stripe is $\partial_t \phi_j$.
The first term, 
$\vec j\cdot
\vec A$ in the  gauge $\vec A = B y \hat x$,
gives rise to the Lorentz force law. 
$\kappa_a$ are the various elastic constants, and $Q$ is the bending stiffness 
of a stripe.  

The term
which tends to  lock the relative phases of the
CDW's on neighboring stripes, ${\cal L}_{lock}$, is the principal term which
distinguishes the crystal from the smectic - it vanishes in a smectic phase.
$\alpha^{-1}$ is the wavelength of the CDW in appropriate units, and  
$\beta= \lambda\alpha$ is necessary to insure
that the system is rotationally invariant\cite{halperin}.  This term
has been omitted in the published literature on this problem, but its necessity can
be seen readily.  Consider a rotated version of the
classical stripe crystal ground state:  $u_j=x\sin(\theta) +j\lambda 
[1-\cos(\theta)]/\cos(\theta)$ and $\phi_j=-j\lambda \sin(\theta)$.  Since this is
also a classical ground-state, it must be that $S=0$.  The effective action we have
considered, with the stated value of $\beta$, is invariant under this transformation for
$\theta$ sufficiently small that  we can ignore
the non-linear terms, {\it i.e.} $[1-\cos(\theta)]/\cos(\theta)\approx 0$.
To insure rotational invariance under large rotations, we would have to include additional
non-linear terms in the fields, as is well known\cite{lubensky} in the classical liquid crystal
literature; however, these terms do not affect any results of the present paper.
C{\^o}t{\'e} and Fertig\cite{fertig-cote} have explicitly shown, in 
Hartree-Fock  approximation, that
$V$ is very small, and indeed, 
we will ignore it for now, and then assess its 
perturbative  relevance, below.

The final term, ${\cal L}_{irr}$, consists of all remaining, higher order
terms which, as we will see, are irrelevant at the quantum Hall smectic fixed
point.  We have explicitly exhibited three of the most interesting of these - 
inertial terms, which can be neglected at low energies and large magnetic
fields, and the leading geometric coupling between the stripe geometry
and the density wave order along the stripe, which plays a key
role\cite{nature,EFKL} in determining the stability 
of the smectic phase in the
absence of a magnetic field.

If we take the continuum limit of this action, 
then the leading order terms in 
${\cal L}_{sm}$ 
have the interpretation as the fixed point action for the quantum Hall smectic.  
This limit is obtained
by replacing $\lambda\sum_j \rightarrow \int dy$, 
$u_j(x)\rightarrow u(x,y)$, $\phi_j(x)\rightarrow \phi(x,y)$, and
${\cal L}_{sm}\rightarrow {\cal L}_{sm}^*$:
\be
{\cal L}_{sm}^*= \frac {eB} {\lambda} u \partial_t \phi - \frac {\kappa_{\parallel}} 2 
(\partial_x\phi)^2 - \frac {\kappa_{\perp}} 2 ( \partial_y u)^2 -\frac Q 2
(\partial_x^2 u)^2.
\label{eq:Lsm}
\ee
In a separate publication\cite{paper2} we give a detailed 
microscopic derivation of this effective action.

This action is a scale invariant fixed point action with respect to the 
anisotropic transformation
\begin{equation}
\begin{array}{ccc}
 x\rightarrow r x &  y \rightarrow r^{2} y & t\rightarrow 
r^3 t  \\
u \rightarrow r^{-1} u & \theta \to \theta & \phi \rightarrow r^{-2} 
\phi
\end{array}
\label{eq:scaling-short}
\end{equation}
where $\theta$ is the dual field defined below, in Eq.\ \ref{eq:dual}.
Thus, the effective space-time dimension is $6$. All the operators included in 
the fixed point
Lagrangian ${\cal L}^*_{sm}$ are marginal since they have scaling dimension $6$. 

If we then perform a standard scaling analysis, all operators with
dimension larger than $6$ are irrelevant while operators with dimension smaller 
than $6$ are relevant. 
It would then be straightforward to assess the perturbative relevance of $S_{lock}$ ;  
in the continuum limit,
we can approximate $1-\cos[\alpha(\phi_j-
\phi_{j+1})+\beta \partial_x u]=(\lambda^2\alpha^2/2)(\partial_y\phi+\partial_xu)^2 + \ldots$,
from which we deduce that 
this term is the energy associated with a shear deformation. By power counting,
at the smectic fixed point this operator is a combination of operators with dimensions $4$, $6$  and
$8$ respectively. The operator $(\partial_xu)^2$ is {\em relevant}, according to this scaling
analysis, suggesting that the smectic is unstable to formation of an unpinned  crystalline state.
The higher order terms,
$\ldots$, can be easily seen to have higher dimension. Similar analysis leads to the conclusion that
the explicit terms in
${\cal L}_{irr}$ have dimensions
$8$ ($M_1$),
$10$ ($M_2$), and $7$ ($\gamma$), respectively.  

It is easy to see that
external periodic potentials ({\it e.\ g.\/} lattice pinning) are 
relevant:  an operator of the form $\cos(2\pi u/\lambda)$ has scaling
dimension $2$ and is strongly relevant,  while operators of the
form $\cos(\beta \phi)$ (where $\beta $ is a constant) have dimension  $4$ and 
are also relevant. Finally, an explicit term of the form $(\partial_x u)^2$ 
(as opposed to the one generated by expanding ${\cal L}_{lock}$) has scaling 
dimension $4$ and is also relevant. Such terms break rotational invariance
explicitly and can be generated for instance by an in-plane magnetic field 
(or by any other perturbation that generates an anisotropic effective mass for
the electrons).

The Lagrangian as written is even under $u_j\rightarrow -u_j$ and
$B\rightarrow -B$ (related to particle-hole symmetry);  
there is another operator\cite{alan} that should appear in ${\cal L}_{sm}$
that we have
not included, namely
$ (u_{j+1}-u_j)\partial_x \phi\rightarrow \lambda(\partial_yu)(\partial_x\phi)$.  However its
effect is equivalent to a renormalization of the elastic constants.
Although by power counting it is marginal, we treat it as a redundant
operator and  will ignore it. 

Another issue, raised originally by MacDonald and Fisher, is whether there are 
two distinct sets of low
energy degrees of freedom or not.  Clearly, since the dynamical term in the 
action linearly couples $u$ and
$\phi$, they are mixed, so there is only one low energy mode, not two.  
Formally, this means that we can
integrate out one set of degrees of freedom, leaving an action with a quadratic 
kinetic energy in terms of
the other.  For example, if we Fourier transform the effective action, and then 
integrate out the shape modes, we are left with
\be
S_{sm}^*=\frac 1 2\sum_{\vec k,\omega}
\frac {\kappa_{\parallel}k_x^2} {\epsilon^2(\vec k)}
\left[ \omega^2 - \epsilon^2(\vec
k)\right] |\phi_{\vec k,\omega}|^2,
\ee
where
\be
\epsilon^2(\vec 
k)=\kappa_{\parallel} \left(\frac{\lambda}{eB}\right)^2 k_x^2
[Qk_x^4 +\kappa_{\perp}k_y^2]
\ee
which is the dispersion relation of the Goldstone modes. This result agrees with the analysis of C\^ot\' e and Fertig~\cite{fertig-cote}.
Notice that it has  a line of values of $\vec
k=(0,k_y)$ with zero energy. 

However, the price we have to pay for integrating out the Goldstone modes $u$,
is a singular dependence of the action on $\omega$ and $\vec k$,
reflecting the presence of highly non-local interactions.  It is a matter of 
convenience, then,
whether we treat the problem in terms of a non-local action with a minimal 
number of degrees of freedom, or
a local action with twice as many degrees of freedom;  we feel that the latter 
representation makes the
basic physics clearer.   In any case, the effective action we obtain here is
equivalent to that of MacDonald and Fisher\cite{MF}. 
In summary, we conclude that the displacement
field $u$, representing the fluctuations of the shape of the stripe, and the Luttinger filed
$\phi$, representing the charge fluctuations on each stripe, are canonically conjugate variables
and thus are not independent degrees of freedom, in agreement with the arguments of
MacDonald and Fisher\cite{MF}. This discussion corrects some of the arguments given earlier 
by two of us in ref.\ \cite{FK}.

Another way to look at the effective action is as a phase-space path integral - 
in
this case, $\phi_j$ and 
\be
\Pi_j\equiv - eB u_j
\ee
are interpreted as a field
and its conjugate momentum, $[\phi_j(x,t),
\Pi_k(x',t)]=i \delta_{jk}\delta(x-x')$. 
(This reflects the well known Landau level physics, in
which the different components of the position operator become canonically
conjugate variables, and so fail to commute.)  We can thus express the same 
physics in terms of a (discretized) Hamiltonian density operator ${\cal
H}=\sum_j {\cal H}_j$ where
\be
{\cal H}_j=\frac{\lambda}{2(eB)^2}[Q(\partial_x^2\hat\Pi_j)^2+\frac
{\kappa_{\perp}} {\lambda^2} (\hat\Pi_j-\hat\Pi_{j+1})^2] +\frac
{\lambda \kappa_{\parallel}}{2} (\partial_x\hat\phi_j)^2
\ee
This formulation is convenient for studying
the effect of single particle  (fermionic)
excitations in the smectic state. 

Standard methods\cite{bosonization} of bosonization for the 1DEG permit us to
reconstruct the electron operators for the
$j$-th stripe from the collective bosonic fields. The operators that  create an
electron on the
$j$-th stripe are related to the left and right moving fields, 
$\psi_{\pm,j}(x,t)$
in  the usual manner
\begin{equation}
\Psi_j(x,t)=e^{ik_F^j x} \psi_{+,j}(x,t)+e^{-ik_F^jx} \psi_{-,j}(x,t)
\label{eq:psi}
\end{equation}
The right and left moving Fermi fields have the bosonized form, in terms of the 
field $\phi_j$ on each stripe, 
\bea
\psi_{\pm,j}(x,t)= {\displaystyle{\frac{{\cal U}_j}{\sqrt{2\pi a}}}} 
e^{\displaystyle{i \sqrt{\pi}[\theta_j(x,t) \pm \phi_j(x,t)]}}
\label{eq:electrons}
\eea
where $a=\ell$ is the short distance cutoff.
Here the dual field $\theta_j$ is defined by
\be
\theta_j(x,t)=\int_{-\infty}^x dx' \Pi_j(x',t)
\label{eq:theta}
\ee
and the operators $U_j$ are the Klein factors
\be
U_j=\prod_{i<j} 
e^{\displaystyle{i \sqrt{\pi}\int_{-\infty}^\infty dx \partial_x \phi_j(x,t))}} 
\label{eq:klein}
\ee
which satisfy the relations 
$U_j^{\dagger}\theta_k(x)U_{j}=\theta_{k}(x)+i\sqrt{\pi}\delta_{jk}$ 
and $[U_{i},U_{j}]=[U_{i},\phi_{j}]=0$. 

It is simple to check that $\Psi_j(x,t)$ constructed in this way satisfies
canonical anticommutation relations for a fermionic field, and adds a charge
$e$ to the system.  Because the bosonized Hamiltonian is quadratic, it is also
straightforward (although a bit complicated) to compute the electron Green 
function;
this is done in Section \ref{sec:CF}.  Remarkably, we find that the fluctuations are so 
severe
that the fermion Green function falls
off as a function of imaginary time like $G(t)\sim \exp[ -A\log^2(t)]$, which is
faster than any power law, although slower than exponential.  As exponential
fall-off implies a gap in the spectrum, this behavior implies a strong pseudo-gap
with a characteristic energy scale.  This is {\em not} the behavior one observes in a
set of coupled one dimensional Luttinger liquids, {\it i.e.} in a pinned smectic.  This is
discussed further in Section
\ref{sec:CF}. 

It is useful to find an effective Lagrangian for the dual field $\theta$. 
This can be
done straightforwardly by noting that the dual field $\theta$ and the 
displacement field $u$ are
related by
\begin{equation}
\partial_x \theta=-\frac{eB}{\lambda} u
\label{eq:dual}
\end{equation}
Thus, upon integrating out the $\phi$ fields in Eq.\ \ref{eq:Lsm}, we find that 
the dynamics of the dual field $\theta$ is governed by the local
effective  Lagrangian
\begin{equation}
{\cal L}[\theta]=\frac{1}{2\kappa_\parallel} \left(\partial_t \theta \right)^2-
\frac{\kappa_\perp \lambda^2}{2e^2B^2}
 \left(\partial_x \partial_y \theta \right)^2-
\frac{Q\lambda^2}{2e^2B^2}  
\left(\partial_x^3  \theta \right)^2
\label{eq:Ltheta}
\end{equation}
Since $\theta$ is dimensionless, all three terms in this effective 
Lagrangian correctly
have scaling dimension six. 
In Fourier space the Lagrangian takes the form
\begin{equation}
S[\theta]=\int \frac{d^3k}{(2\pi)^3} \frac{1}{2\kappa_\parallel}
\left[\omega^2-\epsilon^2(\vec k)\right]|\theta({\vec k},\omega)|^2 .
\end{equation}
This result shows that it is the dual field $\theta$ that can 
most naturally be related with the
collective modes of the quantum Hall smectic phase.

We conclude this section with a few observations concerning the properties of the smectic fixed
point.

{\em Symmetries:}  There is a peculiar ``semi-local'' sliding symmetry of the effective field
theory. Because the Hamiltonian has no terms which depend on the variation of
$\phi$ in the $y$ direction, ${\cal L}_{sm}$ is invariant under the transformation
\be
\phi(x,y,t) \rightarrow \phi(x,y,t) + f(y)
\label{eq:sym}
\ee
where $f(y)$ is an arbitrary function of the (continuum) stripe index $y$. This symmetry corresponds to the
independent translations or shifts along each stripe\cite{OLT}.  Superficially this symmetry is similar to a local gauge symmetry. However, since it is not truly local, as the allowed transformations depend only on the $y$ coordinate, this symmetry can be completely broken by a suitable choice of  boundary conditions, e. g. open boundary conditions (in contrast to a genuine gauge invariance which requires gauge fixing.) There is an analogous sliding symmetry for the $\theta$ fields. Nevertheless, this semi-local sliding symmetry has profound consequences on the behavior of both the $\phi$ and $\theta$ correlation functions which are infrared divergent for $y\neq 0$.
We shall also see that there are a number of striking additional features of the correlation functions which stem from this
symmetry. (The locking term, ${\cal L}_{lock}$, if relevant, breaks this
symmetry.)  Although the smectic state spontaneously breaks translational symmetry
in the $y$ direction, the underlying symmetry of space is reflected in
the invariance of $S$ under
\be
u(x,y,t)\rightarrow u(x,y,t)+ a.
\ee 
Similarly, the underlying rotational symmetry is reflected in the invariance of $S$ under
\begin{eqnarray}
u \to u + \theta x \nonumber \\
\phi \to \phi -\theta y.
\end{eqnarray}
Finally, absent the redundant term $(\partial_x\phi)(\partial_yu)$, the action is invariant
under the particle-hole transformation, $u\rightarrow -u$ and $B\to -B$.

{\em Coulomb interactions:}  In the present discussion, we have assumed all interactions are
short-ranged.  (See Section \ref{sec:CF} for the effects of Coulomb interactions). 

{\em Breakdown of the scaling assumption:}  Finally we note that the scaling analysis implied by
Eq.\
\ref{eq:scaling-short}  assumes that the correlation
functions of the fields $\theta$, $\phi$ and $u$ obey simple albeit anisotropic 
power laws. We will show in the next section that this is (almost) true for the
correlation functions of the displacement fields $u$, but that
the Luttinger field $\phi$ and the
dual field $\theta$ develop logarithmic singularities along the stripe direction. In fact the
correlation functions are not only highly anisotropic but have also a complex singularity
structure, determined by both rotational invariance and by the sliding symmetry. In section
\ref{sec:stability} we will argue what this structure, and in particular the existence of a
divergent number of low energy degrees of freedom, raises questions concerning the validity of
naive scaling, and calls for a more careful renormalization group analysis than we have attempted in
this paper.

\section{Correlation Functions of the quantum Hall smectic}
\label{sec:CF}

In this section, we use the effective field theory to compute the correlation
functions of the
displacement field $u$ (which is the Goldstone boson of the smectic) and of the
Luttinger  
fields $\phi$ and $\theta$, which directly represent the charge fluctuations
on each stripe. We will see  that the correlation functions  can be
written in a scaling form which explicitly displays the scaling laws dictated by
the fixed point.  

We define the imaginary time correlators
\be
C_u(x,y,t) \equiv \langle u(x,y,t)u(0,0,0) \rangle
\ee
(and analogously for $\phi$ and $\theta$), and we will denote
the subtracted correlators by
\be
\tilde C_{\phi}(x,y,t)\equiv -\frac{1}{2} \langle \left[\phi(x,y,t)-\phi(0,0,0)\right]^2 \rangle
\label{eq:subtracted}
\ee
We will find below that due to the smectic symmetry the unsubtracted propagator of 
operators that are not invariant under the shift symmetry of Eq.\ (\ref{eq:sym})
are infrared divergent in the limit $x \to 0$ and $t \to 0$  with $y$ fixed. 
While this fact is true in any theory with gapless excitations, these infrared divergences are
particularly important due to the feature of the dispersion relation $\epsilon(\vec k)$
which vanishes at $k_x=0$ {\sl for all } $k_y$,
Thus, the propagators of
the Luttinger field $\phi$ and of the dual field $\theta$ need to be subtracted.
In contrast, the
propagator of the Goldstone modes $u$ needs no subtraction in the thermodynamic
limit since the $u$ fields are invariant under shifts. Among other
things, this implies that the quantum fluctuations of $u$ are bounded,
and so do not necessarily destroy the smectic order.  However, the semi-local
symmetry of the smectic results in expressions for
$C_{\phi}$ and $C_{\theta}$ which diverge in the thermodynamic limit;  only the
subtracted version of these correlators, which are invariant under this symmetry,
are well defined. Notice, however, that the subtraction of Eq.\ \ref{eq:subtracted}
only removes the uniform shift and not the semi-local shifts. Thus, even the subtracted 
propagators for $\phi$ and $\theta$ will become infrared divergent in the limit $x \to
0$, $t \to 0$, with $y$ fixed. We will deal explicitly with this issue below. 

\subsection{The dual field correlation function}
\label{sec:dual}
We will begin by calculating the subtracted propagator of the dual 
field $\theta$ in imaginary time:
\begin{equation}
{\tilde C_\theta}(x,y,t)=\int \frac{d^2k}{(2\pi)^2} \frac{\kappa_\parallel}{2\epsilon(\vec k)}
\; \left[ e^{\displaystyle{-|t| \epsilon(\vec k)+i \vec k \cdot \vec x}}-1\right]
\label{eq:G-theta}
\end{equation}
In principle this integral must be done with finite ultraviolet cutoffs in both $k_x$
and $k_y$, which we can take to be $\Lambda$, the large momentum cutoff of the edge modes, and
$1/\lambda$, where $\lambda$ is the wavelength of the stripe state, typically a number of the order
of a
few magnetic lengths. Also,
for a finite system of linear size $L_x=L_y=L$, $1/L$ will be the infrared low momentum cutoff. 
However, it turns out that 
this subtracted propagator is finite in the thermodynamic limit $L \to \infty$ except
in the regime $x \to 0$ and $t \to 0$ with $y$ fixed. 
This infrared divergence is a consequence
of the local shift invariance (or sliding symmetry)  of the quantum Hall smectic phase. 
In addition, the subtracted propagator has a finite limit as $\Lambda \to \infty$ and
$\lambda \to 0$, for all space-time points $(x,y,t)$ except on the axes (which will be
discussed below). Thus, for generic values of $(x,y,t)$ the subtracted propagator is
faithfully described by the scaling form,
\begin{equation}
{\tilde C_\theta}(x,y,t)=\frac{1}{2\pi^2}\sqrt{\frac{\kappa_\parallel}{\kappa_\perp}}
\frac{eB}{\lambda} F_\theta(\xi,\eta)
\label{eq:f-theta-scaling}
\end{equation}
where $F_\theta(\xi,\eta)$ is the scaling function
\begin{eqnarray}
F_\theta(\xi,\eta)=\frac{1}{2}  \; \Re \;\int_0^{\infty} &&\frac{du}{u} \int_{-\infty}^\infty
\frac{dz}{\sqrt{1+z^2}}
\nonumber \\
\times
&&
\left[e^{\displaystyle{-\eta u^3 \sqrt{1+z^2}+i\xi u+i u^2 z}}-1\right]
\nonumber \\
&&
\label{eq:F-theta} 
\end{eqnarray}
Here $\xi$ and $\eta$ are the scaling variables
\begin{eqnarray}
\xi&=& \left(\frac{\kappa_\perp}{Q}\right)^{1/4} \; \frac{|x|}{\sqrt{|y|}}
\nonumber \\
\eta&=& \frac{\lambda}{eB} \sqrt{\kappa_\parallel \kappa_\perp}
\left(\frac{\kappa_\perp}{Q}\right)^{1/4}\; \frac{|t|}{|y|^{3/2}}
\nonumber \\
&&
\label{eq:xi-eta}
\end{eqnarray}
Notice that the correlation function ${\tilde C_\theta}(x,y,t)$ depends on its arguments
only through the scaling variables $\xi$ and $\eta$ of Eq.\ \ref{eq:xi-eta}, which are scale
invariant according to the rules of section \ref{sec:phenomenology}, Eq.\
\ref{eq:scaling-short}. This feature
follows from the scaling properties of the dual field $\theta$ which, 
according to Eq.\  \ref{eq:scaling-short}, is invariant under scale transformations.  

It is direct consequence of scaling, and of the arguments presented above,
that $F_\theta(\xi,\eta)$ is a smooth differentiable function for all values 
of the scaling variables $\xi$ and
$\eta$, except close to $(\xi,\eta)=(0,0)$ ( {\it i.\ e.\/} $x \to 0$ and $t \to 0$ 
with $y$ fixed) where the subtracted propagator becomes infrared
divergent, and for $\xi \to \infty$ ( {\it i.\ e.\/} $y \to 0$ with $x$ fixed) 
or $\eta \to \infty$ ( {\it i.\ e.\/} $y \to 0$ with $t$ fixed) where the scaling
function develops branch cut singularities. These branch cuts will show up in the form of
logarithmic dependences on $x$ and $t$ for $y \to 0$. The scaling function
has also an infrared divergence as a function of
$y$ as $x \to 0$ and $t \to 0$. This infrared divergence is a manifestation of the shift
invariance, eq.\ \ref{eq:sym}, of the quantum smectic. 

It would be natural then to further
subtract this (divergent) contribution from the scaling function which would now be infrared
finite everywhere. However, in order to keep things simple, we will refrain from doing this
but we will keep in mind the existence of these divergent contributions.
Thus, we see that for generic values of $\xi$ and $\eta$ the correlation function is a
finite scale-invariant function only of $\xi$ and $\eta$. Notice that fixing both $\xi$ and
$\eta$ generally corresponds to a curve in space and time. 

We will now discuss the three special regimes:\\ 
\noindent
(a) $x=0$, $t$ fixed and $y \to 0$:\\
\noindent
 The auto-correlation function of the dual field $\theta$ has a
strong logarithmic infrared singularity as $(x,y,t) \to (0,0,0)$,
\begin{equation}
C_\theta(0,0,0)=\frac{3}{4\pi^2} \frac{eB}{\lambda}
\sqrt{\frac{\kappa_\parallel}{\kappa_\perp}} \ln^2 \left(\frac{L}{\lambda}\right)
\label{eq:G-theta-t-0}
\end{equation}
This regime corresponds to setting $\xi=0$ and taking $\eta \to \infty$, where the
scaling function $F_\theta$ takes the limiting form
\begin{equation}
F_\theta(0,\eta) \to \ln^2\eta
\label{eq:F-theta-eta}
\end{equation}
Hence, the subtracted correlation function ${\tilde C}_\theta(x,y,t)$ is
infrared finite and has the leading long time behavior
\begin{equation}
{\tilde C}_\theta(0,0,t)=\frac{1}{2\pi^2} \frac{eB}{\lambda}
\sqrt{\frac{\kappa_\parallel}{\kappa_\perp}}
\ln^2\left(\frac{|t|}{t_0}\right) \ldots
\label{eq:G-theta-t}
\end{equation}
where 
\begin{equation}
t_0=\left(\frac{\lambda}{\sqrt{\kappa_\perp}}\right)^{3/2}
\frac{Q^{1/4}}{\sqrt{\kappa_\parallel}} \frac{eB}{\lambda}
\end{equation}
Notice that $t_0 \to 0$ as the UV cutoff in $y$ vanishes, $\lambda \to 0$ 
(naturally, this is done
only inside the cutoff factor which carries the $3/2$ power, and not in the dimensional 
factor $eB/\lambda$.) Thus a finite but small
$y$ acts as a short distance cutoff for the time correlation function.\\
\noindent
(b) $t=0$, $x$ fixed and $y \to 0$:\\
\noindent
In this regime, $\eta=0$ and $\xi \to \infty$, where the scaling function behaves like
\begin{equation}
F_\theta(\xi,0) \to \ln^2\xi
\label{eq:F-theta-xi}
\end{equation}
as $\xi \to \infty$. Hence,
the equal-time subtracted correlation function ``on the same stripe" ({\it i.\ e.\/} 
as $y \to 0$),  has the long distance behavior
\begin{equation}
{\tilde G_{\theta}}(x,0,0)= \frac{1}{2\pi^2} \frac{eB}{\lambda}
\sqrt{\frac{\kappa_\parallel}{\kappa_\perp}} \ln^2\left(\frac{|x|}{x_0}\right) 
\label{eq:G-theta-x}
\end{equation}
where 
\begin{equation}
x_0=\left(\frac{\lambda^2 Q}{4\kappa_\perp}\right)^{1/4}
\end{equation}
Notice that here too  a small but finite $y$ acts as a short distance cutoff for the $x$
correlator.\\
\noindent
(c) $x=t=0$ and $y$ fixed:\\
\noindent
In this regime both $\xi \to 0$ and $\eta \to 0$ and the scaling function develops an
infrared divergence. We find that the equal-time {\sl subtracted} correlation function 
on different
stripes has the behavior
\begin{equation}
{\tilde C_{\theta}}(0,y,0)= \frac{1}{8\pi^2} \frac{eB}{\lambda}
\sqrt{\frac{\kappa_\parallel}{\kappa_\perp}} \ln^2\left(\frac{\pi^2}{2}
\sqrt{\frac{Q}{\kappa_\perp}} \frac{|y|}{L^2}\right) + \ldots
\label{eq:G-theta-y}
\end{equation}
where once again $L$ is the linear size of the system. This correlation function is
also infrared divergent as a consequence of the local sliding symmetry of the $\theta$
field.

\subsection{The displacement field correlation function}
\label{sec:displacement}
We will analyze the propagator of the displacement field along the same lines used above for the
$\theta$ correlator.
The propagator of the displacement field $u$ at space-time separation $x=(x,y,t)$, where
$t$ is {\sl imaginary time}, is given by
\begin{eqnarray}
&&C_u(x,y,t)=\int \frac{d^2k}{(2\pi)^2} 
\; \kappa_\parallel \left(\frac{\lambda}{eB}\right)^2 
\frac{k_x^2}{2\epsilon(\vec k)}  e^{\displaystyle{i\vec k \cdot \vec
r-|t|\epsilon(\vec k)}}
\nonumber \\
&&
\label{eq:Gu}
\end{eqnarray}
The same line of argument used for the $\theta$ correlator now implies that the propagator
of the displacement field $u$ for finite $(x,y,t)$ has the scaling form in the limit of $L
\to \infty$, and, after removing all the ultraviolet cutoffs, we get:
\begin{equation}
C_u(x,y,t)=\frac{1}{2\pi^2} \frac{\lambda}{eB}
\sqrt{\frac{\kappa_\parallel}{Q}} \frac{1}{|y|} F_u(\xi,\eta)
\label{eq:Gu-scaling}
\end{equation}
where $F_u(\xi,\eta)$ is the scaling function 
\begin{eqnarray}
F_u(\xi,\eta)=&&
\frac{1}{2}  \; \Re \;\int_0^{\infty} du \; u \int_{-\infty}^\infty
\frac{dz}{\sqrt{1+z^2}} \nonumber \\
&& \times \;
e^{\displaystyle{-\eta u^3 \sqrt{1+z^2}+i\xi u+i u^2 z}}
\nonumber \\
&&
\label{eq:F-u} 
\end{eqnarray}
which converges provided $\xi$ and $\eta$, defined in Eq.\ \ref{eq:xi-eta}, are finite. 
Hence, consistent with the predictions of the scaling laws of Eq.\ \ref{eq:scaling-short}, 
the full propagator is the
product of the universal scaling function $F_u(\xi,\eta)$ and the power law factor of
$1/|y|$, {\it i.\ e.\/} $u$ scales like $1/r$ and its propagator like $1/r^2$ (where $r$ is
the scale factor).
\\
\noindent
(a) {\sl The auto-correlation function}:\\
In the regime $x \to 0$ and $y \to 0$ with $t$ fixed, {\it i.\ e.\/}  $\xi=0$ and $\eta \to
\infty$, we find that the propagator has the
asymptotic behavior
\begin{equation}
C_u(0,0,t)=-{\displaystyle{\frac{A_t}{|t|^{2/3}}}}
\label{eq:Gu-t2}
\end{equation}
where
\begin{equation}
A_t=\frac{1}{6\pi^2} \left({\displaystyle{\frac{\Gamma(1/3)}{2^{2/3}}}}\right)^2
\left({\displaystyle{\frac{\lambda}{eBQ^4}}}\right)^{1/3}
\end{equation}
This behavior is consistent with the scaling laws.\\
\noindent
(b) {\sl The equal-time correlation function}:\\
The equal-time correlation function $C_u(x,0,0)$ for the displacement 
field has the asymptotic behavior
\begin{equation}
C_u(x,0,0)=-{\displaystyle{\frac{A_x}{|x|^{2}}}} 
\ln \left({\displaystyle{\frac{x_0}{|x|}}}\right)
\label{eq:Gu-x1}
\end{equation}
which is also consistent with scaling. The constant
$A_x$ is given by
\begin{equation}
A_x=\frac{1}{2\pi^2}\sqrt{\frac{\kappa}{\kappa_\perp}}
\frac{\lambda}{eB}
\end{equation}
This result can also be obtained by differentiating twice the scaling function for the
$\theta$ field with respect to $x$, at $\eta=0$.\\
(c) {\sl Equal-time correlation on different stripes}:\\
\noindent
Unlike the $\theta$ field correlation function, discussed in subsection \ref{sec:dual} and
the Luttinger field correlator to be discussed below, the
equal-time correlation function for the displacement field $u$ 
on different stripes, $C_u(0,y,0)$ is infrared finite and has the
expected scaling form 
\begin{equation}
C_u(0,y,0)=-{\displaystyle{\frac{A_y}{|y|}}} 
\label{eq:Gu-y}
\end{equation}
where
\begin{equation}
A_y=\frac{1}{8\pi} \frac{\lambda}{eB} \sqrt{\frac{\kappa_\parallel}{Q}}
\end{equation}
The infrared finiteness of the $u$ correlator is required by the smectic symmetry since the
$u$ fields are the Goldstone bosons of the symmetries broken spontaneously by the quantum
Hall smectic state.\\
\subsection{The Luttinger field correlation function}
\label{sec:luttinger}
We now turn to the (subtracted) propagator of the Luttinger field $\phi$. 
In imaginary time and for finite $(x,y,t)$, the subtracted propagator 
of the Luttinger field is given by 
\begin{equation}
{\tilde C_\phi}(x,y,t)=\int \frac{d^2k}{(2\pi)^2} 
\; \frac{\epsilon(\vec k)}{2 \kappa_\parallel k_x^2}
 \left[ e^{\displaystyle{i\vec k \cdot \vec
r-|t|\epsilon(\vec k)}}-1\right]
\label{eq:G-phi}
\end{equation}
Once again, we will write this propagator in scaling form
\begin{equation}
{\tilde C_\phi}(x,y,t)=\frac{1}{2\pi^2} \sqrt{\frac{\kappa_\perp}{\kappa_\parallel}}
\frac{\lambda}{eB} \frac{1}{y^2} F_\phi(\xi,\eta)
\label{eq:G-phi-scaling}
\end{equation}
which is consistent with scaling. The scaling function $F_\phi(\xi,\eta)$ is
\begin{eqnarray}
F_\phi(\xi,\eta)=\frac{1}{2}  \; \Re \;\int_0^{\infty} &&du \; u^3 \int_{-\infty}^\infty
dz \; \sqrt{1+z^2}\;
\nonumber \\
\times
&&
\left[e^{\displaystyle{-\eta u^3 \sqrt{1+z^2}+i\xi u+i u^2 z}}-1\right]
\nonumber \\
&&
\label{eq:F-phi} 
\end{eqnarray}
This scaling function is well defined for all values of $\xi$ and $\eta$ except near $(0,0)$
(namely, for $x \to 0$ and $t \to 0$ with $y$ fixed)
where it develops infrared singularities, and for $\xi \to \infty$ ($y \to 0$ with $x$
fixed) or $\eta \to \infty$ ($y \to 0$ with $t$ fixed) where
it too develops branch cut singularities. We will consider now these three regimes:\\
\noindent
(a) $x=0$ $y \to 0$ with  $t$ fixed:\\
\noindent
 The auto-correlation function of the Luttinger 
field $\phi$ has a logarithmic infrared divergence as $t \to 0$,
\begin{equation}
C_\phi(0,0,0)=\frac{\Gamma(4/3)}{4\pi^2} \frac{\lambda}{eB}
\left(\frac{\sqrt{\kappa_\perp}}{\lambda^3 \kappa_\parallel}\right)^{1/2}
\ln \left(\frac{L}{x_0}\right)
\end{equation}
where $L$ is the linear size of the system.
In contrast, the subtracted auto-correlation function is infrared finite and has the
asymptotic behavior for $|t| \gg |y|^{3/2}$,
\begin{equation}
{\tilde C_\phi}(0,0,t)=-\frac{\Gamma(4/3)}{3\pi^2} \frac{\lambda}{eB} 
\left(\frac{\sqrt{\kappa_\perp}}{\lambda}\right)^{3/2} \frac{1}{\sqrt{ \kappa_\parallel
\kappa_\perp}} \ln\left|\frac{t}{t_0}\right| 
\label{eq:G-phi-t}
\end{equation}
which seemingly violates scaling. However we will see below that this results also
follows from the scaling function.\\
\noindent 
(b) $t=0$ and $y \to 0$ with $x$ fixed:\\
\noindent
By direct evaluation of the scaling function $F_\phi(\xi,0)$ we find that as $\xi \to
\infty$ it behaves like
\begin{equation}
F_\phi(\xi,0) \to {\rm ci}(\xi)-\ln \xi
\label{eq:F-phi-xi}
\end{equation}
where ${\rm ci}(\xi)$ is the cosine integral. Thus, for large $\xi$ the scaling function has
a logarithmic term similar to the one discussed above. Hence, the equal-time subtracted 
correlation function of the Luttinger 
field $\phi$ has the $|x| \gg \sqrt{y}$ behavior
\begin{eqnarray}
\lefteqn{{\tilde C_\phi}(x,0,0)=-\frac{6}{\pi^2}\frac{1}{\lambda^2} 
\sqrt{\frac{\bar \kappa_\perp}{\bar \kappa_\parallel}}
\frac{\lambda}{eB}
\ln\left|\frac{x}{x_0}\right|
}\nonumber \\
&&-\frac{3}{2\pi^2} \frac{Q}{\sqrt{\kappa_\perp \kappa_\parallel}}
\left(\frac{\lambda}{eB}\right) 
\frac{1}{x^4}+\ldots
\label{eq:G-phi-x}
\end{eqnarray}
Hence we see that, for arbitrary finite values of the scaling variables 
$\xi \sim |x|/\sqrt{|y|}$ 
and $\eta \sim |t|/|y|^{3/2}$, the subtracted propagator of the Luttinger field obeys
strictly the scaling laws, and it develops a logarithmic singularity as either the $t$ or
the $x$ axes are approached.\\
\noindent
(c) $x=0$ and $t=0$ with $y$ fixed:\\
\noindent
 The equal-time correlation function for $\phi$ on different
stripes is also strongly infrared divergent as $L \to \infty$. In particular, since it can
be seen that in this case $\lim_{y \to 0} C_{\phi}(y)=0$, there is no need to subtract the
correlation function. Thus the correlation function is
\begin{equation}
C_\phi(0,y,0)=\frac{1}{2\pi^2} \left(\frac{\lambda}{eB}\right)
\frac{1}{\sqrt{\kappa_\perp \kappa_\parallel}} 
\left(\frac{\lambda}{|y|}\right)^2 
\ln \left[\frac{\lambda L}{x_0 |y|}\right]
\label{eq:G-phi-y}
\end{equation}
which diverges in the thermodynamic limit, $L \to \infty$.
The physical origin of this infrared divergence is, once again, the local shift symmetry of the
Luttinger field on each stripe.

\subsection{The Electron Propagator in the Smectic Phase}
\label{sec:electron}

In this subsection
we will use the fixed point theory to reconstruct the electron operator. 
We are interested in  finding out several things. To begin with, 
we would like to know if the quantum smectic state
does support sharp excitations with the quantum numbers of the electron. At the
Hartree-Fock level there clearly are electron-like excitations in the spectrum. 
However, the
connection between this problem and Luttinger models suggests that it may be 
possible 
to find a behavior akin to an array
of quasi-one-dimensional systems in which the electron 
fractionalizes into a set of suitably defined solitons. In fact, if the 
displacement
fields were to be gapped out, say by lattice commensurability effects, this is 
indeed
what it may well happen as discussed in ref. \cite{EFKL} and 
\cite{carpentier}.
Thus, instead of a Luttinger-like power law behavior, we will find that the
auto-correlation function of the electron, and the associated spectral function, 
vanishes as a function of either $|x|$ or $|t|$ faster than any power but more slowly than
an exponential. This ``pseudo-gap" behavior is indicative of a pronounced 
suppression of final states for electron tunneling in the quantum Hall smectic phase.
 In the remainder of this section we will 
derive this result.

We are interested in computing the Green function of the electron operators as
defined in Section \ref{sec:phenomenology}. In this section we will use the results of section
\ref{sec:CF} to compute the fermion Green function directly in the continuum limit along the $y$
direction. Thus, from now on, we will replace the discrete stripe label $j$ by the continuum
coordinate $y$. Due to the smectic
symmetry\cite{OLT,golubovic,EFKL}, Eq.\ \ref{eq:sym}, of the effective Lagrangian of 
Eq.\ \ref{eq:Lsm}, the (continuum) fermion 
propagator 
\begin{equation}
G(x,y,t;x',y',t')=\langle T \Psi(x,y,t) \Psi^\dagger(x',y',t') \rangle
\end{equation}
vanishes identically for $ y \neq y'$. Thus it is sufficient to compute the 
propagator on a single stripe, {\it i.\ e.\/} as $|y'-y| \to 0$. Below we
will compute the propagator for the right movers. The propagator for the left 
movers follows trivially from it. 

Using the bosonization formulas, the propagator for the right moving fermions  
at fixed $y$ ({\it i.\ e.\/} on the same stripe) $\psi_{+}(x,y,t)$ is
\begin{equation}
\langle \psi_{+}(x,y,t)^\dagger \psi_{+}(0,y,0)\rangle=\frac{1}{2\pi a} e^{-\pi 
\displaystyle{\Phi_{+}(x,t)}}
\end{equation}
where, for  imaginary time $t>0$ and $x>0$, we find
\begin{equation}
\Phi_+(x,t)=\lim_{y \to 0} \left({\tilde C_\theta}(x,y,t)+{\tilde C_\phi}(x,y,t)\right)
\end{equation}
where we have dropped the contributions from the cross correlations between $\theta$ and $\phi$.
The only role of these terms is to insure the correct anticommutation relations on the fermion
operator, {\it i.\ e.\/} that the fermion propagator is an odd, antiperiodic in time, function of
the
coordinates. It is straightforward to check that these conditions are met.

Elsewhere in this section we showed that the correlator of the dual field ${\tilde C_\theta}$ is
always
more singular than the propagator of the Luttinger field ${\tilde C_\phi}$. Using these results
we find that, at equal (imaginary) times and at $y=0$, the function $\Phi_+(x,0,t)$ has the
has the asymptotic behavior for large $|x|$
\begin{equation}
\Phi_{+}(x,0,0)=\frac{1}{2\pi^2}\frac{eB}{\lambda}\sqrt{\displaystyle{\frac{\kappa_\parallel}
{\kappa_\perp}}}
\; \ln^2(\frac{|x|}{x_0}) +
\ldots
\label{eq:Gx0}
\end{equation}
 We can easily see that this asymptotic behavior is due to the
equal-time correlation function of the dual fields, ${\tilde C_\theta}(x)$,
 of Eq.\ \ref{eq:G-theta-x}. Consequently, the
equal-time fermion  correlation  function behaves like
\begin{equation}
G_{+}(x,0,0) \propto {\rm sgn}(x) \; 
e^{\displaystyle{- \frac{1}{2\pi}\frac{eB}{\lambda}
\sqrt{\displaystyle{\frac{\kappa_\parallel}
{\kappa_\perp}}}
\ln^2(\frac{|x|}{x_0})}}
\end{equation}
where the ${\rm sgn}(x)$ factor shows that it is a correlation function 
of a {\sl fermion}. This
correlation function exhibits the same analytic behavior as the structure 
factor in DNA-lipid
complexes\cite{golubovic,OL1}. Notice  that it 
decays {\it faster} than any power but {\it more slowly} that an 
exponential decay.

Conversely, the fermion auto-correlation function (in imaginary time) is, for 
$t>0$, and on the same stripe ($y=0$) is given by
\begin{equation}
G_+(0,0,t) \propto e^{\displaystyle{-\pi \Phi_+(0,0,t)}}
\end{equation}
whose long time behavior, at zero temperature and as finite temperature, 
is once again dominated by ${\tilde C_\theta}(t)$, the
auto-correlation function of the dual field. Hence, we find that the fermion autocorrelation
function
has the long time behavior
\begin{equation}
G_{+}(0,0,t) \propto  \; 
e^{\displaystyle{- \frac{1}{2\pi}\frac{eB}{\lambda}
\sqrt{\displaystyle{\frac{\kappa_\parallel}
{\kappa_\perp}}}
\ln^2(|t|/t_0)}}
\label{eq:fermion-t}
\end{equation}
with the same $t_0$ defined above in this section. 

The fermion Green function on different stripes can be found by similar means. Using these
methods, the equal-time Green function on different stripes, $y \neq 0$, is found to vanish,
\begin{equation}
G_+(0,y,0) \propto e^{\displaystyle{-\pi \Phi_+(0,y,0)}} \to 0
\end{equation}
in the thermodynamic limit $L \to \infty$ due to the infrared divergence in both ${\tilde
C_\theta}(0,y,0)$ and in ${\tilde C_\phi}(0,y,0)$, which are both a consequence of the sliding
symmetry of the quantum Hall smectic state.
 
The behavior of the  electron auto-correlation function of Eq.\ \ref{eq:fermion-t}
clearly shows the strong suppression of electron states at low energies. 
It implies a  dramatic suppression of the {\sl tunneling density of states} for
electrons into the quantum Hall smectic.
Notice that this is
a much more dramatic effect than the power law behavior of one-dimensional Luttinger
liquids. It is due to the strong fluctuations of the shape of the stripes, a hallmark of
a quantum smectic\cite{nature}. It easy to see that in the presence of
a periodic pinning potential along the direction perpendicular to the stripes, this
behavior is superseded by a conventional Luttinger picture. This is expected since the
resulting state is equivalent to the smectic metal of ref. \cite{EFKL}. 
Similarly, it is
easy to see that at conventional Luttinger behavior is recovered at short times once
irrelevant operators which lead to a linear dispersion relation are included.

\section{The Coulomb Fixed Point}
\label{sec:coulomb}

In the previous sections we have discussed in detail the properties of the quantum
Hall smectic phase for a 2DEG with short range interactions. Here we will consider
the effects of long range Coulomb interactions. Using a time-dependent
Hartree-Fock approach C{\^ o}t{\' e} and Fertig\cite{fertig-cote} have
investigated the quantum Hall smectic state in a 2DEG with Coulomb interactions,
and found that the dispersion relation of the collective modes is modified. In
this section we will investigate how these changes modify the picture of the
fixed point developed above.
 
Rather than reworking the mean field theory we have used for the short range case,
we will work directly with the effective Lagrangian of Eq.\ \ref{eq:Lsm} and
modify it to reflect the effects of Coulomb interactions. There is a simple and
direct way to account for their effects at the level of the effective theory.
First of all, the term $\frac{1}{2} \kappa_\parallel \left( \partial_x
\phi\right)^2$  represents short-range density-density interactions (recall that
in bosonization the forward scattering part of the electron local density is
given by $\partial_x \phi$). Thus long-range interactions lead to a non-local term
in the {\sl action}. Let ${\tilde V}(\vec k)$ be the Fourier transform of the
two-dimensional Coulomb interaction, {\it i.\ e.\/} 
${\tilde V}(\vec k)\propto 1/|\vec k|$.
In momentum space this amounts to modify
$\kappa_\parallel$ by a momentum-dependent factor
\begin{equation}
\kappa_\parallel \to \kappa_\parallel(\vec k)=\frac{\bar \kappa_\parallel}{|\vec k|}
\label{eq:kappa-parallel-coulomb}
\end{equation}
where ${\bar \kappa}$ is an effective coupling constant. Similarly, long-range
Coulomb interactions will also modify the elastic modulus $\kappa_\perp$ along the direction
perpendicular to the stripes. Hence, $\kappa_\perp$ will also have to be changed
in a similar fashion,
\begin{equation}
\kappa_\perp \to \kappa_\perp(\vec k)=\frac{\bar \kappa_\perp}{|\vec k|}
\label{eq:kappa-perp-coulomb}
\end{equation}
Thus the effective low energy theory of the quantum Hall smectic with Coulomb
interactions is obtained by replacing $\kappa_\parallel \to \kappa_\parallel(\vec
k)$ and $\kappa_\perp \to \kappa_\perp(\vec k)$ at the level of the effective
action, in Fourier space. The new dispersion relation is
\begin{equation}
\epsilon(\vec k)= \frac{\lambda}{eB} \; \sqrt{\frac{\bar \kappa_\parallel}{|\vec k|}}
\; |k_x| \; \sqrt{\displaystyle{Q k_x^4+\frac{\bar \kappa_\perp}{|\vec k|} k_y^2}}
\label{eq:dispersion-coulomb}
\end{equation}
Thus, the dispersion relation is non-local. However we will show below that for the analysis of
the infrared behavior of the displacement fields correlators it is sufficient to use an 
approximate, simpler, version of the dispersion relation
obtained by
setting $|\vec k| \approx |k_x|$. In this limit, the dispersion relation 
becomes
\begin{equation}
\epsilon(\vec k)\equiv 
\frac{\lambda}{eB} \; \sqrt{\bar \kappa_\parallel} 
\sqrt{\displaystyle{Q |k_x|^5+{\bar \kappa_\perp} k_y^2}}
\label{eq:dispersion-coulomb-2}
\end{equation}
In particular, at $k_y=0$, we find $\epsilon(k_x,0)\propto |k_x|^{5/2}$, 
a result
first obtained by C{\^ o}t{\' e} and Fertig\cite{fertig-cote}. 

The dispersion relation of Eq.\ \ref{eq:dispersion-coulomb-2} shows that the
dimensional counting for the case of  Coulomb interactions is $\omega^2\sim k_y^2
\sim k_x^5$, which suggests new scaling transformations, with scale factor $r$,
 of the form
\begin{equation}
\begin{array}{ccc}
x \to r x &
y \to r^{5/2}  y
&
t \to r^{5/2}  t
\\
u \to r^{-1} u
&
\phi \to r^{-5/2} \phi
&
\theta \to \theta
\end{array}
\label{eq:scaling-coulomb}
\end{equation}
Hence, in the case of Coulomb interactions the effective dimension is
$D=1+5/2+5/2=6$. Note that the same caveats that were raised about the short range scaling
laws also apply here.

However, unlike the short range case, for $k_x=0$ this approximate dispersion 
relation now vanishes only at $k_y=0$, rather than on the entire line $k_x=0$ as it is 
the case
for the exact dispersion relation as a consequence of the sliding symmetry. 
However, this is not a problem for the calculation of the correlator
of the displacement field since this field is invariant under the sliding symmetry. For the
same reason this approximation cannot be used to calculate the corraltors of the Luttinger and
the dual fields since these are not invariant under the sliding symmetry. 

We have computed explicitly the behavior of the
low energy and long distance behavior of the correlation functions. We found the following
 results:
\subsection{The dual field propagator}

We computed the subtracted correlator of the dual field $\theta$ for the Coulomb case. 
For $x$, $y$ and $t$ close to the axes we find:
\begin{enumerate}
\item
For $t=y=0$, the
(subtracted) correlation function has the behavior
\begin{equation}
{\tilde C_\theta}(x,0,0)=\frac{5}{8\pi^2} \frac{eB}{\lambda} \sqrt{\frac{\bar
\kappa_\parallel}{\bar \kappa_\perp}} 
\ln^2\left((4{\bar \kappa_\perp}/Q\lambda^2)^{1/5}|x|\right)
\end{equation}
\item
In the regime $x=0$, $y\to 0$ and $t$ finite, we get
\begin{eqnarray}
{\tilde C_\theta}(0,0,t)&=&-\frac{2}{5\pi^2} \frac{eB}{\lambda}\sqrt{\frac{\bar 
\kappa_\parallel}{\bar \kappa_\perp}} \ln^2 \left(t \frac{\lambda}{eB} \sqrt{{\bar 
\kappa_\parallel}{\bar \kappa_\perp}} \Lambda \right) 
\nonumber \\
&&
\end{eqnarray}
where $\Lambda$ is the UV cutoffs, which we took to be the same along the x and y 
directions for simplicity. Here we have kept 
only the $\log^2 t$ terms.
 Thus ${\tilde C_\theta}(t)$ and ${\tilde C_\theta}(x)$ have a similar infrared behavior.
\item
In contrast, for $x=t=0$ we find instead the infrared divergent behavior
\begin{equation}
{\tilde C_\theta}(0,y,0)=\frac{1}{16\pi^2} \frac{eB}{\lambda} \sqrt{\frac{\bar
\kappa_\parallel}{\bar \kappa_\perp}} 
\ln^2 \left(\left(\frac{\pi}{L}\right)^{5/2} \sqrt{\frac{Q}{\bar \kappa_\perp}}
|y|\right)
\end{equation}
similar to what we found for the case of short range interactions.
\end{enumerate}
\subsection{The propagator of the displacement field}
We verified that for the calculation of the infrared behavior of the 
correlation functions of the displacement field it
is correct to replace the full dispersion by the approximate one.

we find that the propagator of the displacement field $u$ takes the scaling form
\begin{equation}
{\tilde C_u}(x,y,t)=\frac{1}{2\pi^2} \frac{\lambda}{eB}
\sqrt{\frac{{\bar \kappa_\parallel}}{{\bar \kappa_\perp}}} 
\left(\frac{\bar \kappa_\perp}{Q}\right)^{2/5}
\frac{1}{|y|^{4/5}}
F_u^c(\xi,\eta)
\label{eq:G-u-scaling-coulomb}
\end{equation}
where $F_u^c(\xi,\eta)$ is the scaling function 
\begin{eqnarray}
F_u^c(\xi,\eta)=\frac{1}{2}  \; \Re &&\int_0^{\infty} du \; u \int_{-\infty}^\infty
\frac{dz}{\sqrt{1+z^2}}
\nonumber \\
\times
&&
e^{\displaystyle{-\eta u^{5/2} \sqrt{1+z^2}+i\xi u+i u^{5/2} z}}
\nonumber \\
&&
\label{eq:F-u-coulomb} 
\end{eqnarray}
In limiting cases we find the asymptotic behaviors
\begin{eqnarray}
C_u(x,0,0)&=&-A_x
\frac{\ln (\mu |x|)}{|x|^2}
\nonumber \\
C_u(0,y,0)&=&\frac{A_y}{|y|^{5/2}}
\nonumber \\
C_u(0,0,t)&=&\frac{A_t}{|t|^{5/2}}
\nonumber \\
&&
\label{eq:Gu-coulomb}
\end{eqnarray}
which, up to logarithms, clearly obey the scaling laws of eq.\ \ref{eq:scaling-coulomb}.
Here we have set
\begin{eqnarray}
A_x&=&\frac{5}{8\pi^2} \frac{\lambda}{eB} \sqrt{\frac{\bar
\kappa_\parallel}{\bar \kappa_\perp}}
\nonumber \\
\mu&=&(4 {\bar \kappa_\perp}/Q)^{1/5}
\nonumber \\
A_y&=&\frac{(\sqrt{5}-1)}{20\pi^2} \Gamma(2/5) \Gamma(1/10) \Gamma(4/5) 
\frac{{\bar \kappa_\parallel}\left(\lambda/eB\right)^2}{\left({\bar \kappa_\perp}Q^4\right)^{2/10}}
\nonumber \\
A_t&=&\frac{\left(\Gamma(2/5)\right)^2}{5\pi^2} 
\left(\frac{\lambda}{eB}\right)
\left(\frac{\bar
\kappa_\parallel}{\bar \kappa_\perp}\right)^{1/10}
\left(\frac{eB}{2\sqrt{2} \lambda Q}\right)^{4/5} 
\nonumber \\
&&
\end{eqnarray}
Hence, even in the presence of Coulomb interactions the propagator of the 
displacement field $u$ obeys scaling. Notice that here too this propagator is free of the
infrared singularities that we find in both the $\theta$ and in the $\phi$ correlators.
Once again this feature is dictated by the smectic symmetry.

\subsection{The propagator for the Luttinger field}

The subtracted propagator of the Luttinger field does not have a simple scaling form 
in the Coulomb case.

We will only  consider the limiting regimes:
\begin{enumerate}
\item
For $t=0$ and $y=0$, the subtracted correlation function becomes
\begin{equation}
{\tilde C_\phi}(x,0,0)=-
\frac{1}{8\pi^2}\frac{1}{\lambda^2}\sqrt{\frac{\bar \kappa_\perp}{\bar
\kappa_\parallel}} \frac{\lambda}{eB}\ln\left(\frac{|x|}{x_0}\right)+\ldots
\end{equation}
where $x_0=(Q/{\bar \kappa_\perp})^{1/5} \lambda^{2/5}$. This behavior is similar to what
we found in the short range case.
\item
For $x=y=0$, we find that the subtracted correlator has the behavior 
\begin{equation}
{\tilde C_\phi}(0,0,t)=-
\frac{\lambda}{eB} \sqrt{\frac{\bar 
\kappa_\perp}{\kappa_\parallel}} \frac{1}{4\pi^2} 
\Lambda^2 \ln \left(t \frac{\lambda}{eB} 
\sqrt{{\bar
\kappa_\parallel}{\bar \kappa_\perp}} \Lambda \right) + \ldots
\end{equation} 
where $\Lambda=\Lambda-x=\Lambda_y$ is the UV cutoff.
Thus, the subtracted autocorrelation function has a $\ln t$ behavior. 
\item
For $\eta=0$ and $ \xi \to 0$, we now find
\begin{equation}
{\tilde C_\phi}(0,y,0)=-\frac{5}{8\pi^2} \frac{\lambda}{eB} \sqrt{\frac{\bar
\kappa_\perp}{\bar \kappa_\parallel}} \frac{1}{y^2} \ln \left(\left(Q \lambda^2/{\bar
\kappa_\perp}\right)^{1/5} L \right)
\end{equation}
Hence, here too this propagator is infrared divergent.
\item
The strong multiplicative infrared divergences that we found in 
$C_\phi(t)$ imply that, for Coulomb interactions too,
the fermion propagator vanishes for $y\neq 0$. However, just as in the case of short range
interactions, at $y=0$ the fermion propagator is dominated by the contribution of the
correlator of the dual field $\theta$, which here too behaves as $\log^2x$ or $\log^2 t$.
Hence, also in the case of Coulomb interactions, the fermion autocorrelation function has
the same ``pseudogap" behavior found for the case of short range interactions, {\it i.\ e.\/}
$\exp(-A \log^2 (t/t_0))$ (where $A$ is a constant).
\end{enumerate}

\section{Thermodynamic Properties of the Quantum Hall Smectic}
\label{sec:thermo}

The quantum Hall smectic phase has remarkable thermodynamic properties. These can be deduced
directly form
the effective low energy theory. 

It is an elementary exercise in statistical mechanics to show that the internal energy density
$U$ is 
\begin{equation}
U=\int \frac{d^2k}{4\pi^2}  \frac{\epsilon(\vec k)}{e^{\epsilon(\vec k)/T}-1}
\label{eq:U}
\end{equation}
We will apply this formula for both cases, short range and Coulomb interactions. For 
short range interactions we find that the internal energy density at low temperatures has the
behavior
\begin{equation}
U(T)=\frac{1}{9}\frac{\lambda}{eB} {\sqrt{\frac{\kappa_\parallel}{ \kappa_\perp}}} T^2
\ln\left(\frac{T_0}{T}\right)+\ldots
\label{eq:Ushort}
\end{equation}
where $T_0=1/t_0$.
Hence, the low temperature specific heat $c(T)$ of the quantum Hall smectic phase obeys the law
\begin{equation}
c(T) \propto T \ln \frac{T_0}{T}+ O(T, T^3 \ln T)
\label{eq:c-short}
\end{equation}
Hence, for short range interactions, the low temperature specific heat of the 2DEG in the quantum
Hall smectic phase is larger than the specific heat of either Fermi or Luttinger liquid phases
(both being linear).

Coulomb interactions modify the above laws only slightly. 
In particular we find that the internal energy density at low
temperatures now obeys a pure $T^2$ law (without the logarithm), and that the specific heat has
a $T$ linear behavior, $c(T) \sim T$.

\section{Is the Quantum Hall Smectic Phase Stable?}
\label{sec:stability}

We have shown that, although the Goldstone modes of the smectic are very soft, the
quantum zero-point fluctuations of the long-wave-length modes do not destroy the
stripe order.  This is not a trivial issue - at finite temperature, the harmonic analysis
would lead to a linearly diverging mean-square stripe displacement, $< u^2 > \sim T L$.
Of course, what this really means is that there is no smectic phase at finite temperatures.

It is a much more subtle issue  whether the smectic phase is unstable to the formation of a
stripe crystal, {\it i.e.} whether there is  inevitably translation symmetry breaking along the
stripe direction.
There are two sorts of analysis that have been applied to answer this question,
although we feel there are possible flaws with both.

In Section \ref{sec:phenomenology} we defined a scale invariant smectic fixed point Lagrangian. 
Treating this Lagrangian as one would in studying critical phenomena, one can assess the
relevance of various physically allowed perturbations using standard dimensional analysis.  By
this analysis, there is one operator, which originates from the shear piece of the locking term in
the stripe crystal Lagrangian, which is apparently relevant.  One might conclude from this that the
smectic is unstable to crystallization. A bizarre aspect of this analysis, which causes us to
have reservations concerning its validity, is that the different pieces of the locking term,
whose relative strength is ultimately determined by rotational symmetry, have different scaling
dimensions. 
This analysis was generalized for the case of Coulomb interactions in
Section
\ref{sec:coulomb}; the stability analysis yields the same answer.
The basic
assumption behind the scaling analysis is that the correlation functions 
have a simple analytic structure reflecting the scaling laws, although $x$, $y$, and $t$ 
enter with different effective exponents, in just the same way that space and time can enter
differently in quantum critical phenomena. 

However, in Sections \ref{sec:CF} and \ref{sec:coulomb} we found that while the Goldstone boson
field $u$ follows the scaling laws (up to a multiplicative logarithmic correction), the Luttinger
field $\phi$ and the dual field
$\theta$ have a much more complex singularity structure. In particular the correlation functions of
the fields $\phi$ and $\theta$ for space-time points on the same stripe (that is, for equal $y$
coordinates) have the leading equal-time behavior $\log (|x|/x_0)$ and $\log^2(|x|/x_0)$ respectively 
(and an analogous behavior as a function of time). Along other directions the $\phi$ and $\theta$ are 
finite only when properly subtracted. Otherwise they exhibit infrared divergences, as a
consequence of the sliding symmetry. The consequent breakdown of scaling is
``weak''  - it only involves logarithms.  However, the RG analysis, although based on examining lowest order
perturbation theory in the additional couplings, is ultimately non-perturbative;  it is based on the
assumption that the correlation functions scale.  Without a careful analysis of the singularity
structures that occur in higher order perturbation theory, it is not possible to
determine whether the breakdown of scaling is significant or not.  Indeed, 
the length and time scales $x_0$ and $t_0$ discussed in Section
\ref{sec:CF} appear in various correlation functions - for instance,
the fermion
propagator has a ``pseudo-gap" behavior
$\exp(-{\rm const.}~
\log^2(|t|/t_0))$, instead of the familiar power law of Luttinger liquids.  
These properties seemingly imply that, unlike arrays of Luttinger
liquids, the QH smectic is not a critical state. Instead it behaves much more like a stable state of
matter, characterized by the finite length and time scales. 

The second approach is based on the observation that the on-stripe pure logarithmic
behavior of the $\phi$ correlator is reminiscent of that encountered in weakly coupled Luttinger
liquids. This singularity has been used to suggest that the locking perturbations may ultimately
drive the QH smectic state to a crystalline state \cite{FK,MF,bert}.  Moreover, if the
structure of lowest order perturbation theory  in the ${\cal L}_{lock}$ is examined, it is found to
be dominated by large values of $k_y$, so the characteristic two dimensional dispersion of the
smectic Goldstone modes does not significantly affect the results.

When the stripes are pinned ({\it i.e.} if the transverse Goldstone behavior is suppressed), we
have no doubt of the validity of this approach. 
In such a state the Goldstone bosons are gapped and can be
integrated out. Their net effect is to induce and renormalize the inter and intra-stripe forward
scattering interactions. Hence, the pinned smectic reduces to an array of Luttinger liquids. 
On a technical level, this approach actually treats the system as a distinct Luttinger liquid for each value
of $k_y$.  Consequently, instead of a Luttinger parameter and Fermi velocity, there is a ``Luttinger function''
$K(k_y)$ and a velocity function $v_F(k_y)$.  The phase
diagram of such systems was considered recently by us \cite{EFKL}, by Vishwanath and
Carpentier\cite{carpentier}, and by Sondhi and Yang \cite{sondhi-yang}.
It was found that, as the
parameters of the Luttinger function are changed, there is a complex phase diagram which includes both
crystalline and smectic metallic phases. 

However,  there is reason to worry about the validity of this  approach for the unpinned smectic.
Specifically, this analysis is completely  insensitive to the small $k_y$ behavior of the Luttinger and
velocity functions.  But the Goldstone behavior of the smectic implies a vanishing Fermi velocity as $k_y\to
0$.  This pathology has no consequences in lowest order perturbation theory, which is why it does not affect
the scaling analysis that is based on it.  However, it is responsible for the the form of the low
temperature specific heat and the non-Luttinger liquid (pseudo-gap) form of the  fermion propagator in the
quantum smectic state.  It seems too good to be true that this salient physics 
should have no affect on the stability of the state.  Again, it is only by an analysis of the structure of
higher order perturbation theory (which we do not attempt) that it can be determined whether the pathologies
associated with small
$k_y$ are truly unimportant for this purpose, or whether they become more important in higher order
terms. 

Leaving this issue aside, we can approach the stability question from the viewpoint of coupled Luttinger
liquids.
MacDonald and Fisher
\cite{MF} showed that rotational invariance constrains the Luttinger functions to reach simple limiting
values as a function of $k_y$. They further argued that if these
functions are monotonically increasing with $k_y$, and if charge-conjugation symmetry is
respected, then the CDW lock-in interactions are always relevant and the QH smectic is ultimately unstable to
crystallization.  This is in apparent contradiction with the results obtained by  Fertig and coworkers
\cite{fertig,fertig-cote,fertig-stability} (using the same method of
analysis)  in which numerical solutions of the time-dependent Hartree-Fock equations
were used to compute the values of $K(k_y)$. 
It is important to remember, in this case, that the scaling analysis performed in Section I can not be used
to discard a variety of additional operators from the fixed-point Hamiltonian.  In particular, there is an
infinite number of marginal operators, representing forward scattering density-density interactions, which
affect the scaling properties\cite{EFKL}.  
The analysis of MacDonald and Fisher \cite{MF} could in principle
be reconciled with these results if the assumption of monotonicity of the Luttinger function on the
transverse momentum is dropped. It is apparent that these neglected terms could, in principal, do this.


\section{Conclusions and Open Problems}
\label{conclusions}

In this paper we presented an effective theory for the low energy degrees of
freedom of the quantum smectic or stripe phase of the 2DEG in a large 
magnetic field.
In the quantum Hall smectic phase the 2DEG spontaneously  breaks both
translational (in the direction perpendicular to the stripes) and
rotational invariance. We showed that the form of the effective theory is
dictated by very general principles: the native symmetries of the smectic
state and the Lorentz force law. 
We determined the spectrum of collective 
modes which exhaust the low energy degrees of freedom. For a 2DEG with short range
(screened) interaction, these Goldstone modes have a dispersion which is
exceedingly soft, vanishing like $\epsilon \sim k_x^3$ at $k_y=0$ and with a whole line of
zero-energy states at $k_x=0$ and $k_y\neq 0$. These Goldstone bosons 
should be detectable in Raman scattering experiments in 2DEG in
heterostructures.  A tangible reflection of how soft
these Goldstone modes are is seen from the temperature dependence of
the specific heat $c_V\sim T|\log(T/T_0)|$ at low $T$.  This is larger than that of
a Fermi liquid! In contrast, we found that for long range Coulomb interactions 
the
Goldstone bosons behave like $\epsilon \sim k_x^{5/2}$ at $k_y=0$. 
Nevertheless, the specific heat obeys the familiar $T$ linear law.

We also used the fixed point theory to compute the electron propagator.  
We found that for short range interactions there should be a pseudo-gap in the single 
particle density of states, as deduced from the tunneling conductance at low temperatures. 
For Coulomb interactions we found a similar behavior. 

The question of the stability of the stripe state has been a matter of discussion and controversy
for some time\cite{nature,FK,MF}. 
We hav shown here that although it is true that by a
suitable parametrization of the degrees of freedom it is possible to write the effective
theory in terms of what looks like an array of coupled Luttinger liquids, this is not a true
quasi-one-dimensional system. Instead, it is a strongly anisotropic, fully two-dimensional
system whose the symmetries force an anisotropic form of
scaling described in Section \ref{sec:phenomenology}. 

Nevertheless, clearly the quantum Hall smectic has very unusual scaling properties.
For instance, although the correlation functions obey scaling, as
shown in great detail in Sections \ref{sec:CF} and \ref{sec:coulomb}, the scaling functions
are not analytic everywhere in the plane defined by the scaling variables $\xi$ and $\eta$,
and develop singularities in the extreme regimes $\xi \to \infty$ or $\eta \to \infty$. However,
these are very soft, logarithmic, integrable singularities on a set of measure zero of
the space-time coordinates. Thus, when the effects of any
typical perturbation is considered, for instance the coupling of the CDW order parameters on
different stripes, it may well be that these additional singularities may not affect the scaling analysis
as they do not lead to true singular behavior of a true two-dimensional system. In any case a more careful
renormalization group analysis is required to reach a definitive conclusion.
However, if either
translation invariance is broken explicitly (say by an external periodic potential in the
direction perpendicular to the stripes) the behavior will indeed cross over to a
quasi-one-dimensional regime quite similar to what was discussed in ref\ \cite{EFKL}. In
other words, a pinning potential changes the properties of this state substantially 
and in fact
it enhances the effects of fluctuations whose main effect is to make it more unstable to 
crystallization. It has also been suggested recently\cite{sondhi-yang} 
that the pinned stripe state
may also be unstable to a quantum Hall state.

An open and very interesting question is the possible existence of a 
quantum nematic state of the 2DEG in large magnetic fields at zero temperature.
This is an important question both conceptually and experimentally as it appears 
to be consistent with the experimental data\cite{FKMN}. Recently some of 
us\cite{vadim} developed a theory of a quantum 
nematic Fermi fluid at zero external magnetic field in the proximity of an
isotropic Fermi liquid phase. It will be particularly interesting to construct a
theory of the quantum melting of the smectic by a
dislocation unbinding mechanism. (While this paper was being refereed Radzihovsky and Dorsey posted a paper on a theory of the quantum Hall nematic\cite{RD})

It is also intriguing to explore the relation of the quantum Hall smectic with
other phases of the quantum Hall system.  There is some evidence from
exact diagonalization studies in small systems\cite{HR,rezayi} of a direct
transition to a 
paired quantum Hall state\cite{read-moore,gww}.  We would expect such a
transition, if it occurs, to be first order.  More exciting is the possibility
of a relation between the quantum Hall smectic and the other well known
compressible state of a half-filled Landau level. 

Finally it is interesting to note that the imaginary time form of the 
effective action for the dual field $\theta$, Eq.\ \ref{eq:Ltheta}, bears a
close formal resemblance to the free energy 
for the sliding columnar phase of 
DNA-lipid complexes\cite{golubovic,OL1,OL2,GLO}. In momentum space the free
energy of the sliding columnar (SC) phase is given by\cite{GLO}
\be
F_{\rm SC}=\int \frac{d^3q}{(2\pi)^3} \left[B q_z^2+K q_x^4+K_y q_x^2
q_y^2\right]\left|u_z(\vec q)\right|^2
\label{eq:FSC}
\ee
where $\hat y$ is the direction normal to the layers, $\hat z$ is the direction
normal to the DNA strands, $u_z$ is the displacement field of the DNA strands
parallel to the
lipid layers, and $B$, $K$ and $K_y$ are effective elastic constants. 
Although
the effective elastic theory are quite similar, the coordinates $x$, $y$, and
$z$ do not scale like the coordinates $x$, $y$ and $t$ of the quantum smectic.
While in the case of the quantum hall smectic the effective dimension is
$D=6$, in the sliding columnar phase it is $D=5$. A direct consequence of these
scaling properties is that
the non-linear corrections to the elastic strain tensor introduce
additional interactions which are marginally relevant in the sliding columnar
phase\cite{OL2}, but are irrelevant in the quantum Hall
smectic phase.
Remarkably, although the details of the models
are different, some of the correlation functions in the quantum Hall
smectic phase  that we discussed in Sections
\ref{sec:CF} and \ref{sec:coulomb} have the same behavior as certain correlation 
functions in the sliding columnar phases. 

\section{Acknowledgments}

We thank A.\ Dorsey, H.\ Fertig, M.\ P.\ A.\ Fisher, T.\ C.\ Lubensky and  A.\ H.
MacDonald for useful discussions, and in particular for helping us to achieve our current state
of confusion concerning the stability issue.  EF thanks J.\ Cardy for useful comments on RG in
anisotropic theories. We also thank Bert Halperin for an insightful exchange about rotational
symmetry and for bringing to our attention the importance of the shear strain
$\partial_y \phi +\partial_x  u$ in a rotationally invariant system.
 This work was supported in part by grants of the National
Science Foundation numbers DMR98-08685 (SAK) and DMR98-17941 (EF). D.\ G.\ B. was 
partially supported by  the University of the State of Rio de Janeiro, Brazil 
and by the Brazilian agency CNPq through a postdoctoral fellowship.




\begin{thebibliography}{99}


\bibitem{Lilly}
M.\ P.\  Lilly, K.\  B.\  Cooper, J.\  P.\  Eisenstein, L.\  N.\  Pfeiffer,
and K.\  W.\  West, 
Phys.\  Rev.\  Lett.\ {\bf 82}, 394 (1999).
\label{ref:Lilly}

\bibitem{du}
R.\ R.\ Du, D.\ C.\ Tsui, H.\ L.\ Stormer, L.\ N.\ Pfeiffer, K.\ W.\ Baldwin KW, 
and K.\
W.\ West, Solid State Comm.\ , {\bf 109}, 389 (1999).
\label{ref:du}

\bibitem{subsequent} 
W.\ Pan, J.\ S.\ Xia, V.\ Shvarts, E.\ D.\ Adams, R.\ R.\ Du, H.\ L.\ Stormer, 
D.\ C.\ Tsui, L.\ N.\ Pfeiffer, K.\ W.\ Baldwin
and K.\ W.\  West, Physica A {\bf 6}, 14 (2000);
M.\ P.\  Lilly, K.\  B.\  Cooper, J.\  P.\  Eisenstein, L.\  N.\  Pfeiffer,
and K.\  W.\  West, 
Phys.\  Rev.\  Lett.\ {\bf 83}, 824 (1999);
W.\ Pan, R.\ R.\ Du, H.\ L.\ Stormer, D.\ C.\ Tsui, L.\ N.\ Pfeiffer, K.\ W.\
Baldwin, and K.\ W.\ West, 
Phys.\ Rev.\ Lett.\ {\bf 83}, 820 (1999);
K.\ B.\ Cooper, M.\ P.\ Lilly, J.\ P.\ Eisenstein, L.\ N.\ Pfeiffer and K.\ W.\ 
West,
Phys.\ Rev.\ B {\bf 60} R11285 (1999).
\label{ref:subsequent}

\bibitem{platzman}
H.\ Fukuyama, P.\ Platzman and P.\ W.\ Anderson, Phys.\ Rev.\ B {\bf 19},
5211 (1979).
\label{ref:platzman}

\bibitem{fogler}
A.\ A.\ Koulakov, M.\ M.\ Fogler, and B.\ I.\ Shklovskii,
Phys.\ Rev.\ Lett. {\bf 76}, 499 (1996);
M.\ M.\ Fogler, A.\ A.\ Koulakov abd B.\ I.\ Shklovskii,
Phys.\ Rev.\ B {\bf 54}, 1853 (1996); M.\ M.\ Fogler and A.\ A.\ Koulakov,
Phys.\ Rev.\ B {\bf 55}, 9326 (1997).
\label{ref:fogler}

\bibitem{chalker}
R.\ Moessner and T.\ J.\ Chalker, Phys.\ Rev.\ B {\bf 54}, 5006 (1996).
\label{ref:chalker}

\bibitem{tudor}
T.\ Stanescu, I.\ Martin and P.\ Phillips, Phys.\ Rev.\ Lett. {\bf 84 }, 1288 
(2000).
\label{ref:tudor}

\bibitem{nature}
S.\ A.\ Kivelson,  E.\ Fradkin and V.\ J.\ Emery,
Nature {\bf 393}, 550 (1998).
\label{ref:nature}

\bibitem{FK}
Eduardo Fradkin and  Steven A.\ Kivelson, Phys.\ Rev.\ B {\bf 59}, 8065 (1999).
\label{ref:FK}

\bibitem{laughlin}
R.\ B.\ Laughlin, Phys.\ Rev.\ Lett.\ {\bf 50}, 1395 (1983).
\label{ref:laughlin}

\bibitem{fertig}
H.\ Fertig, Phys.\ Rev.\ Lett.\  {\bf 82}, 3693 (1999)
\label{ref:fertig}

\bibitem{MF}
 A.\ H.\ MacDonald and Matthew P.\ A.\ Fisher, Phys.\ Rev.\ B {\bf 61}, 5724 
(2000).
\label{ref:MF}

\bibitem{FKMN}
Eduardo Fradkin, Steven A.\ Kivelson, Efstratios Manousakis and Kwangsik Nho, 
Phys.\  Rev.\  Lett.\ {\bf 84}, 1982 (2000).
\label{ref:FKMN}

\bibitem{nematic}  
C.\ Wexler and A.\ Dorsey, Phys.\ Rev.\ B {\bf 64}, 115312 (2001).
\label{ref:nematic} 

\bibitem{vadim}
Vadim Oganesyan, Steven Kivelson, and Eduardo Fradkin,
Phys.\ Rev.\ B {\bf 64}, 195109 (2001).
\label{ref:vadim} 

\bibitem{paper2}
Daniel Barci and Eduardo Fradkin, 
{\sl Theory of the quantum Hall smectic II}, cond-mat/0106171.
\label{ref:paper2}

\bibitem{deGennes} 
P.\ G.\ de Gennes and J.\ Prost, ``The physics of Liquid Crystals'', 
Oxford University Press, New York, 1998. 
\label{ref:deGennes}

\bibitem{lubensky}
P.\ M.\ Chaikin and T.\ C.\ Lubensky,
{\sl Principles of Condensed Matter Physics}, Cambridge University Press 
(Cambridge, UK)
\label{ref:lubensky}

\bibitem{EFKL}
V.\ J.\ Emery, E.\ Fradkin,  S.\ A.\ Kivelson, and T.\ C.\ Lubensky, 
Phys.\ Rev.\ Lett.\ {\bf 85}, 2160 (2000).
\label{ref:EFKL}

\bibitem{fertig-cote}
R.\ C\^ot\'e and H.\ A.\ Fertig, 
Phys.\ Rev.\ B {\bf 62}, 1993 (2000).
\label{ref:fertig-cote}

\bibitem{fertig-stability}
Hangmo Yi, H.\ A.\ Fertig and  R.\ C\^ot\'e, 
Phys.\ Rev.\ Lett.\  {\bf 85}, 4156 (2000).
\label{ref:fertig-stability}

\bibitem{fogler-vinokur}
M.\ Fogler and V.\ Vinokur, Phys.\ Rev.\ Lett.\ {\bf 84}, 5828 (2000).
\label{ref:fogler-vinokur}

\bibitem{bert}
Anna Lopatnikova, Steven H.\ Simon, Bertrand I.\ Halperin and Xiao-Gang Wen,
{\sl Striped states in the quantum Hall effect: deriving a low energy theory
from Hartree-Fock}, cond-mat/0105079.
\label{ref:bert}

\bibitem{halperin}
We are grateful to Bert Halperin for bringing to our attention the linear combination
$\partial_y \phi +\partial_x  u$, representing shear strain, the only combination allowed in a
rotationally
invariant system.

\bibitem{alan}
We thank Alan Dorsey (private communication)
for raising the question of the role of this operator to us.

\bibitem{bosonization}
See for instance V.\ J.\ Emery in 
{\sl Highly Conducting One-Dimensional Solids}, J.\ Devreese {it et.\ al.\/} 
editors,
Plenum (New York, 1979); {\sl Bosonization} M.\ Stone editor, World Scientific 
(Singapore,
1994).
\label{ref:bosonization}

\bibitem{OLT}
C.\ S.\ O'Hern, T.\ C.\ Lubensky and J.\ Toner,
Phys.\ Rev.\ Lett.\ {\bf 83}, 2745 (1999).
\label{ref:OLT}

\bibitem{carpentier}
Ashvin Vishwanath and David Carpentier, 
Phys.\ Rev.\ Lett.\ {\bf 86}, 676 (2001).
\label{ref:carpentier} 

\bibitem{golubovic}
L.\ Golubovi{\' c} and M.\ Golubovi{\' c}, Phys.\ Rev.\ Lett.\ {\bf 80}, 4341 
(1998); Erratum,
Phys.\ Rev.\ Lett.\ {\bf 81}, 5704 (1998). 
\label{golubovic}

\bibitem{OL1}
C.\ S.\ O'Hern and T.\ C.\ Lubensky,
Phys.\ Rev.\ Lett.\ {\bf 80}, 4345 (1998).
\label{OL1}

\bibitem{sondhi-yang}
S.\ L.\ Sondhi and Kun Yang, Phys.\ Rev.\  B {\bf 63}, 054430 (2001) .
\label{ref:sondhi-yang}

\bibitem{RD}
L.\ Radzihovsky and A.\ Dorsey, {\sl Theory of Quantum Hall Nematics}, cond-mat/0110083.
\label{ref:RD}

\bibitem{HR}
Exact diagonalization of 2DEGs with small numbers of electrons suggest this 
possibility;
see  E.\ H.\ Rezayi,  F.\ D.\ M.\ Haldane and Kun Yang, 
Phys.\ Rev.\ Lett.\ {\bf 85}, 5396 (2000).
\label{ref:HR}

\bibitem{rezayi}
E.\ H.\ Rezayi, F.\ D.\ M.\ Haldane and Kun Yang,  Phys.\ Rev.\ Lett.\ {\bf 83}, 
1219(1999).
\label{ref:rezayi}

\bibitem{read-moore}
N.\ Read and G.\ Moore, Nucl.\ Phys.\ B {\bf 360}, 362 (1991).
\label{ref:read-moore}

\bibitem{gww}
M.\ Greiter, X.\ G.\ Wen and F.\ Wilczek, Phys.\ Rev.\ B {\bf 46}, 9586 (1992).
\label{ref:gww}

\bibitem{OL2}
C.\ S.\ O'Hern and T.\ C.\ Lubensky,
Phys.\ Rev.\ E {\bf 58}, 5948 (1998).
\label{OL2}

\bibitem{GLO}
L.\ Golubovi{\' c}, T.\ C.\ Lubensky and C.\ S.\ O'Hern, 
Phys.\ Rev.\ E {\bf 62}, 1069 (2000).

\end{thebibliography}
\end{document}